# Short Paper: Static and Microarchitectural ML-Based Approaches For Detecting Spectre Vulnerabilities and Attacks

Chidera Biringa
cbiringa@umassd.edu
University of Massachusetts
Dartmouth, USA

Gaspard Baye
bgaspard@umassd.edu
University of Massachusetts
Dartmouth, USA

Gökhan Kul
gkul@umassd.edu
University of Massachusetts
Dartmouth, USA

## ABSTRACT
Spectre intrusions exploit speculative execution design vulnerabilities in modern processors. The attacks violate the principles of isolation in programs to gain unauthorized private user information. Current state-of-the-art detection techniques utilize microarchitectural features or vulnerable speculative code to detect these threats. However, these techniques are insufficient as Spectre attacks have proven to be more stealthy with recently discovered variants that bypass current mitigation mechanisms. Side-channels generate distinct patterns in processor cache, and sensitive information leakage is dependent on source code vulnerable to Spectre attacks, where an adversary uses these vulnerabilities, such as branch prediction, which causes a data breach. Previous studies predominantly approach the detection of Spectre attacks using the microarchitectural analysis, a reactive approach. Hence, in this paper, we present the first comprehensive evaluation of static and microarchitectural analysis-assisted machine learning approaches to detect Spectre vulnerable code snippets (preventive) and Spectre attacks (reactive). We evaluate the performance trade-offs in employing classifiers for detecting Spectre vulnerabilities and attacks.

### ACM Reference Format:
Chidera Biringa, Gaspard Baye, and Gökhan Kul. 2022. Short Paper: Static and Microarchitectural ML-Based Approaches For Detecting Spectre Vulnerabilities and Attacks. In . , 5 pages.

## 1 INTRODUCTION
Speculative execution [29] is a $\mu$arch method used to improve modern microprocessor performance. In 2018, Kocher *et al.* [18] showed that components that support speculative execution of assembly instructions such as branch predictions leave quantifiable side effects in processor caches along with other shared resources even with the absence of instruction commit. Spectre attacks [18] are a class of $\mu$arch attacks that pose a significant threat to a computer's security by revealing private user data through a side-channel cache-timing attack. Spectre-variant attacks exploit processor branch prediction to obtain the victim's data. A holistic and aggressive fix implies rethinking the contract between the instruction set architecture (ISA) and $\mu$arch [12]. This realization led to the development of several detection [11, 20, 26, 34] and mitigation [36] solutions. Currently, $\mu$arch-based detection techniques are dependent on hardware performance measures [4, 20, 21], which dictate the distribution of cache stress levels – hits or misses over time. However, attacks can still propagate with adversarial manipulation of the performance counters profiling tools such as `perf`. On the other hand, detection using Spectre vulnerable code snippets [26, 34] is constrained to known vulnerabilities and side-channel data breaches. Several Spectre mitigation strategies such as LFENCE [14] and Kernel Page Table Isolation (KPTI) [24] for solving Meltdown attacks [22] attempt to obtain a viable concession between performance and security. This approach institutes a strictly enforced security that invariably leads to a sub-optimal processor performance [23, 27]. Meltdown is an attack variant similar in principle to Spectre that exploits pipelined access to memory during out-of-order execution to compromise and leak user data from the kernel mode of the CPU. Recent studies in the literature [4, 21, 34, 38] have approached the problem of Spectre attacks as a learning problem with malicious and benign classes and consequently applying Machine Learning (ML) classifiers to detect this attacks using either vulnerable victim programs or hardware performance counters (HPCs).

In this study, we are motivated by: (i) the recent availability of a significant volume of Spectre gadgets to perform Spectre vulnerabilities detection using ML [34], and (ii) a gap in a thorough presentation of the trade-offs in performance between ML classifiers using vulnerabilities and attacks data. Before 2021, conducting ML-assisted Spectre vulnerabilities experiments via victim program was constrained to no more than 17 observations [11, 17], which is not ideal and attributed to the fact that ML classifiers, especially neural networks require large volumes of data to be sufficiently trained and explored [7]. Tol *et al.* [34] solved this problem by using a combination of mutational fuzzing and deep learning (DL) to generate a significant number of Spectre-V1 gadgets suitable for ML and DL experiments. We approach detecting Spectre attacks from both signature vulnerabilities in code and behavioral characteristics in CPU-Processes State (CPS) by leveraging the traces of malicious activity caused by Spectre attacks in the $\mu$arch and detecting Spectre vulnerabilities using gadgets. We propose a comprehensive performance evaluation of static and cache analysis-assisted machine learning approaches to detect Spectre-vulnerable programs and attacks.

**Outline.** The rest of this paper is organized as follows. In §2, we present the necessary background for this work, §3 briefly discusses related work. Our methodology is described in §4 and evaluated in §5. Finally, §6 concludes this paper.

## 2 BACKGROUND

**Speculative Execution.** Speculative execution is an optimization mechanism with significant performance advantages [16]. The primary goal is to reduce the latency of instructions by utilizing idle hardware resources while preventing hazards caused by changes in control order that stall instructions.

**Static Analysis.** Static analysis is the automated analysis of source code without its execution [1]. We statically analyze gadgets, which are victim code vulnerable to Spectre attacks. Although we focus on Spectre V1, vectors generated during analysis can be easily applied to other attack variants. Gadgets facilitate the leaking of data using instruction speculation. There are two main and known types of vulnerabilities [18] which include: (i) Vulnerability via Victim Code Exploits (VCE) and (ii) Vulnerability via Return Stack Buffer (RSB) and Branch Target Buffer (BTB) exploits. Our focus in this paper is VCE, because VCE is the most common type of Spectre attack through manipulation of speculative code snippet [21].

**$\mu$arch Analysis.** $\mu$arch analysis is the exploration of a microprocessor's functionalities [32]. Researchers have shown the importance of comprehensively analyzing the $\mu$arch to detect side-channel attacks [32]. We leverage this knowledge to inspect modern CPU special-purpose registers referred as Hardware Performance Counters (HPC). The HPC is applied to count the occurrences of different CPU event types, such as CPU clock cycles, independent cache level cache hits, and cache misses.

**Spectre Attack Variants.** Spectre variants include: (i) variants 1, CVE-2017-5753 (bounds check bypass on loads) [16], (ii) 1.1, CVE-2018-3693 (bounds check bypass on stores) [16], (iii) 1.2 (read-only protection bypass) [16], (iv) 2, CVE-2017-5715 (branch target injection) [25], (v) 3, CVE-2017-5754 (rogue data cache load) [25], (vi) 3a, CVE-2018-3640 (rogue system register read) [16], (vii) 4, CVE-2018-3639 (speculative store bypass) [16], and (viii) CVE-2017-5715 (branch history injection) [19]. In this paper, **CVE-2017-5753 is our focus point**.

**Spectre Attack Model.** In our attack model, we assume the role of an adversary with the capacity to cause a data breach in the CPU cache through vulnerable victim functions. Listing 1 is a sample example from the original Spectre paper [18]. In the given case, the adversary anticipates the branch condition to return `true`, the code is called with an index outside the given `array1` [16], and the processor speculatively executes `array1[x]` which makes x vulnerable. The code execution traces still reside in the cache because even though the processor finds the branch condition is `false`, it does not scrub the cache. The adversary proceeds to launch a cache timing side-channel attack on the cache (e.g., prime and probe [31]) to uncover `[array1[x] * 512]`. Prime and probe side-channel is a self-contained attack and requires no additional information other than the gadget. An adversary primes `temp &= array2[array1[x] * 512]` from caching the data before CPU memory access guarantees the availability of target data. Next, it probes the cache to obtain `array2` memory access timing data and ensures the change in access points using 512 increments. To complete the attack, the adversary will yield control to a gadget with the capacity to leak private user information via a side channel.

```
void victim_function_v01(size_t x) {
    if (x < array1_size)
        temp &= array2[array1[x] * 512];
}
```

**Listing 1: Exploiting Speculative Execution To Conduct Bounds Check Bypass on Loads Attack.**

## 3 RELATED WORK

Several research work have been proposed for detecting Spectre vulnerabilities [26, 30, 34] and attacks [4, 11, 13, 21, 35, 35, 38]. FastSpec [34] merges mutational fuzzing — a testing method to identify vulnerabilities that lead to software crashes, memory leaks by feeding it randomized inputs [33], and generative adversarial networks (GANs) — a subset of generative modeling using DL, first proposed by Goodfellow [10] — methods to detect Spectre vulnerabilities. SpecFuzz [26] detects Spectre-V1 vulnerabilities using fuzzing to introduce speculative exposure, necessitating the isolated execution of speculation in instructions to expose vulnerabilities. Detection tools such as SPECTECTOR [11], KLEESpectre [35], and SpecuSym [13] utilize symbolic execution to detect speculative vulnerabilities in assembly instructions. oo7 [36], and SpecTaint [30] employs taint analysis to investigate binary files of vulnerable code snippets that result in data breaches. Although the aforementioned research studies propose approaches for detecting Spectre vulnerabilities or attacks. However, to the best of our knowledge, our work is **the first to conduct a comprehensive performance assessment of ML approaches to detect vulnerabilities and attacks facilitated by static and $\mu$arch analysis**.

## 4 METHODOLOGY

**Overall Strategy.** Our methodology as shown in Figure 1 is summarily decomposed into four main phases: (i) in §4.1, we collect benign and malicious Spectre data through performing (a) offline static analysis on gadgets and (b) online $\mu$arch analysis on CPS, (ii) using observations derived from the previous phase, we select and extract CPS and gadgets features, (iii) in §4.2, we feed vulnerabilities and attacks observations to machine learning classifiers, and (iv) finally, in §5 we comprehensive assess the performance of our classifiers in predicting Spectre vulnerabilities and attacks.

### 4.1 Data Collection and Feature Extraction

**Extracting Gadgets Features Using Static Code Analysis.** We statically analyzed Spectre gadgets [34]. Victim (victim functions vulnerable to attacks) and non-victim (disassembled Linux libraries) functions are encoded {1} and {0} respectively. These gadgets in their raw form are C source code transformed to assembly instructions. An indispensable component of building predictive models is to ensure that observations are represented numerically [7]. ML or DL models such as CNN compute using vectors [28]. Thus, we solicit the use of word2vec embedding model [9] — a means of representing words in high-dimensional vectors in relatively low-dimensional vector space — to represent our corpora. We treat independent instructions and functions as words and documents. We describe our analysis and feature extraction in the following steps: (i) removing non-representative and irrelevant information such as directory paths, and file format from our corpora, (ii) tokenizing observations — entails the splitting of text into independent blocks [7], (iii) setting the sequence maximum length (maxlen) to 256, observations ⪕ maxlen is zero-padded and truncated respectively [5]. Currently, there isn't a consensus in the literature on what constitutes a suitable embedding size, only mostly task-dependent recommendations [7]. However, after experimenting with embeddings of varying sizes, we found that a 256 length with 32-dimensional vectors suffices in maintaining a permissible relationship between classification performance and training time in our case. Finally, (iv) we train our model and extract the resulting embeddings as gadget features.

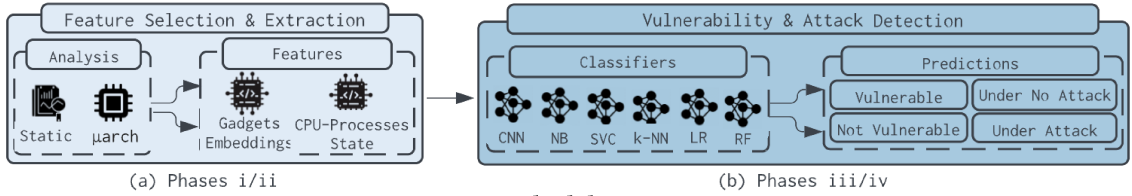

Figure 1: Methodology Overview.

**Extracting CPU-Processes State (CPS) Features Using $\mu$arch Analysis.** Modern CPUs have special-purpose registers called Hardware Performance Counters (HPC). These registers count the occurrences of different CPU event types, such as CPU clock cycles, independent cache level cache hits, and cache misses. In this section, we detail the technique and tools used to collect the CPS data. Both `PAPI` (Performance Application Programming Interface) and `Perf` are widely used command-line tools to collect, visualize, filter, and aggregate data accumulated through HPCs [2]. `perf-stat`, a part of `Perf`, monitors specific CPU events for a given process or thread. The shortest interval on `Perf-stat` between two consecutive samples is 100ms [3]. On the other hand, `PAPI` shows a maximum resolution of 3 μs; hence more than 3000 times faster than `Perf-stat` [2]. This difference is not crucial for performance profiling but is useful in detecting side-channel attacks because it provides fine-grained information on cache activity. The CPS is produced from HPC from each process rather than using the accumulated readings from the CPU. The per-process extraction adjusts the required granularity for our system to quickly notify the user about a suspicious process. In our experiments, we collected data points based on three processor events, which are the L3 cache accesses (L3_TCA), L3 cache misses (L3_TCM), and the total number of instructions (TOT_INST).

Previous work [18, 22, 37] have shown that Spectre uses cache-side channel attacks to exfiltrate data. Cache side-channels like the Flush+Reload repeatedly flush specific fragments of memory from the cache and calculate the access times of a memory read process. Therefore, during a cache side-channel attack, the adversary activity shows higher cache misses. Thus, L3 cache misses (L3_TCM) are a fitting indicator to detect cache side channels. An adversary can stage a cache side-channel attack using specific instructions such as `CFLUSH`. Attackers can intentionally evict the cache using these instructions, which indicates inspecting the L3 cache can identify such acts. L3 cache hits usually correlate with their counterpart cache misses. Therefore, we collect this information to create a relation with cache misses, enabling a comprehensive view of cache activities. The total number of instructions (TOT_INS) captures the complete workload on the CPU. During a cache-side channel attack, we observe that the TOT_INS is related to the number of cache misses due to a malicious process that repeatedly attacks the victim until it succeeds. Consequently, we can deduce that percentage of a cache miss is highly related to the total number of instructions rather than legitimate processes. We use a popular baseline dataset produced by [4] to generate a more diverse and realistic dataset based on our analysis of the $\mu$arch. Having a diverse dataset aims to create a more realistic general-purpose classification model. We label the data points from benign processes as benign {B} and encoded {0}, and those from Spectre processes are labeled malicious {M} and encoded {1}.

### 4.2 Datasets and Classifiers

**Datasets Statistics.** In this study, we collect ~5000 observations of benign and malicious examples of Spectre gadgets and CPS data. The distribution of the gadgets data is 60-40 percent (%) for malicious and benign observations. The CPS data is 1/5 malicious, and the rest benign. Each observation recorded at least 555 processes per service. We account for this data imbalance in our experimental evaluation in §5. Based on the pedagogical literature regarding training and testing sets [8], we split and shuffle our vulnerabilities and attacks datasets into 80-20%, where 80% denotes (70-10%) for training and validations sets, and 20% for our testing set. Furthermore, we used a k-fold cross-validation technique [7] to evaluate our training data for ten iterations.

**Implementing Classifiers.** We implemented five ML classifiers to detect Spectre vulnerabilities and attacks. Table 2 presents selected classifiers. To derive a representative collection of predictive models, we select our classifiers based on deep (CNN), information (RF), probabilistic (NB), non-probabilistic (SVC), and error-based (LR) learning approaches [15]. To maintain experimentation fairness, we kept our classifiers as basic as possible, refraining from using any advanced feature engineering or hyperparameter tuning that might skew the results in favor of one classifier over the other [7].

## 5 EXPERIMENTAL EVALUATION

**Setup.** We conducted experiments using an Apple commodity laptop with an Apple M1 8-core CPU, 16-core Neural Engine, 8 GB RAM, macOS Big Sur (version 11), and a 2.6 GHz Intel Core i7-6700HQ processor, 16 GB RAM, Ubuntu (version 20.04.1) LTS.

**Reproducibility.** We have made our source code and datasets publicly available: https://github.com/biringaChi/SPECDET.

**Metrics.** We experimentally evaluate our models using classical ML classification metrics [7].

**Performance (P).** To comprehensively measure the performance of our classifiers, we employ F-measure, Fbeta-measure, precision, recall, specificity, and geometric-mean metrics [7]. F-measure, recall, and geometric-mean are used to account for an imbalance in data.

$$P_1 = \frac{2 * P * R}{P + R} \quad P_2 = \frac{((1 + \beta^2) * P * R)}{\beta^2 * P + R} \quad P_3 = \frac{\mathbb{R}_{s \to s}}{\mathbb{R}_{s \to s} + \mathbb{R}_{b \to s}}$$

Table 1: A Taxonomy of ML Classification Models for Detecting Spectre Vulnerabilities and Attacks.

| Class | Classifier | Short Description |
|---|---|---|
| Neural Network | Convolutional Neural Network (1D-CNN) | Specialized design of a NN for processing data represented using matrix structures |
| Bayesian Network | Naive Bayes (NB) | Bayesian classifier based on the Bayes' theorem |
| Support Vector Machines | Support Vector Classifier (SVC) | Linear model that employs a hyperplane to maximize class margins |
| Regression Analysis | Logistic Regression (LR) | Employs a probabilistic estimation between dependent and independent variables |
| Decision Tress | Random Forest (RF) | Ensemble of independently trained decision trees also known as estimators |

Table 2: Performance comparison between classifiers on detecting Spectre vulnerabilities and attacks. Clf: Classifier, $F_1$: F-measure, F-b: Fbeta-measure, Pre: Precision, Rec: Recall, Spec: Specificity, G-M: Geometric Mean, Avg: Average, TRT: Time it takes to train and cross-validate data for ten folds. PRT: Time it takes to predict 1000 observations using the trained model.

| | Spectre Vulnerabilities | | | | | | | | | | | | | | | |
|---|---|---|---|---|---|---|---|---|---|---|---|---|---|---|---|---|
| | Training (Cross-Validation($\overline{k=10}$)) | | | | | | | | Testing | | | | | | | |
| Clf. | $F_1$ | F-b. | Pre. | Rec. | Spec. | G-M. | Avg. | TRT(s) | $F_1$ | F-b. | Pre. | Rec. | Spec. | G-M | Avg. | PRT(s) |
| 1D-CNN | N/A | N/A | N/A | N/A | N/A | N/A | N/A | 10.363 | 0.964 | 0.979 | 0.989 | 0.940 | 0.984 | 0.962 | 0.969 | 0.176 |
| NB | 0.970 | 0.986 | 0.998 | 0.944 | 0.997 | 0.970 | 0.977 | 0.328 | 0.972 | 0.985 | 0.994 | 0.951 | 0.992 | 0.971 | 0.977 | 0.043 |
| SVC | 0.999 | 0.998 | 0.998 | 1.000 | 0.997 | 0.998 | 0.998 | 3.964 | 0.998 | 0.997 | 0.996 | 1.000 | 0.995 | 0.997 | 0.997 | 0.804 |
| LR | 0.998 | 0.998 | 0.997 | 1.000 | 0.996 | 0.998 | 0.997 | 1.202 | 0.999 | 0.998 | 0.998 | 1.000 | 0.997 | 0.998 | 0.998 | 0.013 |
| RF | 0.998 | 0.999 | 1.000 | 0.997 | 1.000 | 0.998 | 0.998 | 4.352 | 1.000 | 0.999 | 0.999 | 0.999 | 1.000 | 0.999 | 0.999 | 0.105 |
| | Spectre Attacks | | | | | | | | | | | | | | | |
| 1D-CNN | N/A | N/A | N/A | N/A | N/A | N/A | N/A | 3.126 | 0.000 | 0.000 | 0.000 | 0.000 | 0.995 | 0.000 | 0.165 | 0.053 |
| NB | 0.602 | 0.487 | 0.431 | 1.000 | 0.777 | 0.881 | 0.696 | 0.009 | 0.601 | 0.485 | 0.430 | 1.000 | 0.775 | 0.880 | 0.695 | 0.015 |
| SVC | 0.998 | 0.997 | 0.996 | 1.000 | 0.999 | 0.999 | 0.969 | 14.037 | 0.999 | 0.983 | 0.979 | 1.000 | 0.996 | 0.998 | 0.992 | 0.048 |
| LR | 0.889 | 0.866 | 0.851 | 0.934 | 0.971 | 0.952 | 0.9105 | 8.295 | 0.897 | 0.875 | 0.860 | 0.937 | 0.974 | 0.955 | 0.916 | 0.003 |
| RF | 0.999 | 0.999 | 1.000 | 0.998 | 1.000 | 0.999 | 0.999 | 0.559 | 1.000 | 0.999 | 1.000 | 1.000 | 0.999 | 1.000 | 0.999 | 0.034 |

$$P_4 = \frac{\mathbb{R}_{s \to s}}{\mathbb{R}_{s \to s} + \mathbb{R}_{s \to b}} \quad P_5 = \frac{\mathbb{R}_{b \to b}}{\mathbb{R}_{b \to s} + \mathbb{R}_{b \to b}} \quad P_6 = \sqrt{R * S}$$

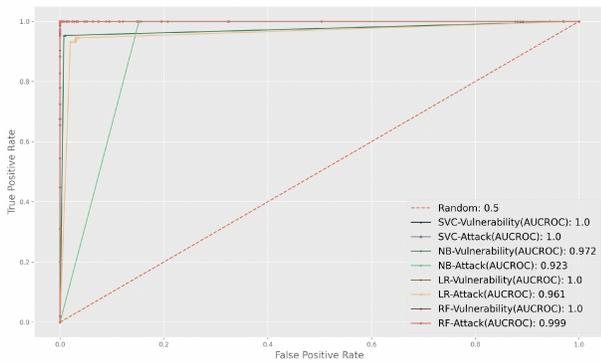

Figure 2: Receiver Operating Characteristic (ROC) Curve.

$P_1$: **F-measure.** Denoted by Equation $P_1$, it calculates the harmonic precision and recall average performance of our models, where $P$ is precision and $R$ is recall.

$P_2$: **Fbeta-Measure.** Denoted by Equation $P_2$, is a derivation of F-measure, where $\beta$ is used to regulate the precision and recall stability during the computation of the harmonic mean.

$P_3$: **Precision (P).** Denoted by Equation $P_3$ measures the ratio of accurately classifying Spectre against the total number of positive Spectre vulnerabilities or attacks, where $\mathbb{R}$ represents the number of observations in the dataset, $(s \to s)$ is predicting Spectre as Spectre, and $(b \to s)$ is predicting benign activity as Spectre.

$P_4$: **Recall (R).** Denoted by Equation $P_4$ is also known as Sensitivity, it calculates the ratio of True Spectre activity classified as Spectre, where $(s \to b)$ is miss-classifying Spectre attacks activity as benign.

$P_5$: **Specificity (S).** Denoted by Equation $P_5$, defines the true negative rate of correctly predicting the negative class, where $(b \to b)$ is predicting benign observations as benign.

$P_6$: **Geometric-Mean.** Denoted by Equation $P_6$ stabilizes the geometric considerations between specificity and recall.

$P_7$: **Area Under the Receiver Operating Characteristic Curve (AUCROC).** A diagnostic tool to visualize the classifiers' capacity to discriminate between classes, where ROC plots (recall) against (1 - specificity). AUC denotes the area under the ROC curve [7].

**Result Discussion.** Highlighted grey in the table cells is the best performing classifier for the corresponding performance metric. On average, SVC and RF perform the best across metrics on detecting Spectre vulnerabilities and attacks for cross-validation and testing sets. On training speed, 1D-CNN produced an outlier performance and is attributed to the embedding size (256x32) dimensional vectors. We can solve this problem by reducing embedding size without a loss in information to maintain an acceptable performance overhead. Figure 2 visualize the comparative ROC curve performance of all classifiers.

# 6 CONCLUSION

**Threats to Validity.** Limitations of our work include: (i) semi-synthetically generated gadgets, (ii) adversarial attacks on classifiers, (iii) lack of granular and ablation analysis on classifiers to empirically ascertain why certain models perform better than others, and (iv) detection focus on Spectre-V1. The aforementioned threats will be addressed in our future work.

**Final Thoughts.** Considering the high semantic gap between a system's compiler and µarch [6], Spectre attack detection is particularly challenging to accomplish since ISA shields software from irregularities occurring in the hardware. Hence, we have presented a comprehensive performance evaluation of proactive (vulnerability) and reactive (attack) approaches to tackle the problem of Spectre using machine learning. Furthermore, we recommended the development and adoption of preventive measures towards tackling Spectre as recent variants have proved to be more evasive evidenced by the recent discovery of Spectre-BHB [19].

**Acknowledgments.** This work has been funded by UMass Dartmouth Cybersecurity Center. Usual disclaimers apply. We want to thank Adnan El-Nasan, Ph.D., for his instruction in CIS 570 (Advanced Computer Systems) course, where we conceived the original idea of this work.